# ENSURING QoS GUARANTEES IN A HYBRID OCS/OBS NETWORK

Sunish Kumar O S

Asst. Professor, Department of Electronics and Communication Engineering, Amaljyothi College of Engineering, Kerala, India
sunishkumaros@gmail.com

## ABSTRACT

*The bursting aggregation assembly in edge nodes is one of the key technologies in OBS (Optical Burst Switching) network, which has a direct impact on flow characteristics and packet loss rate. An optical burst assembly technique supporting QoS is presented through this paper, which can automatically adjust the threshold along with the increasing and decreasing volume of business, reduce the operational burst, and generate corresponding BDP (Burst Data Packet) and BCP (Burst Control Packet). In addition to the burst aggregation technique a packet recovery technique by restoration method is also described. The data packet loss due to the physical optical link failure is not currently included in the QoS descriptions. This link failure is also a severe problem which reduces the data throughput of the transmitter node. A mechanism for data recovery from this link failure is vital for guaranteeing the QoS demanded by each user. So this paper will also discusses a specific protocol for reducing the packet loss by utilizing the features of both optical circuit switching (OCS) and Optical Burst switching (OBS) techniques.*

## 1. INTRODUCTION

As we begin the new millennium, we are seeing dramatic changes in the telecommunications industry that have far-reaching implications for our lifestyles. There are many drivers for these changes. First and foremost is the continuing, relentless need for more capacity in the network. This demand is fueled by many factors. The tremendous growth of the Internet and the WorldWideWeb, both in terms of number of users and the amount of time, and thus bandwidth taken by each user, is a major factor. Internet traffic has been growing rapidly for many years. Estimates of growth have varied considerably over the years, with some early growth estimates showing a doubling every four to six months. Despite the variations, these growth estimates are always high, with more recent estimates at about *50%* annually.

QoS refers to the capability of a network to provide better service to selected network traffic over various technologies, including Frame Relay, Asynchronous Transfer Mode (ATM), Ethernet and 802.1 networks, SONET, and IP-routed networks that may use any or all of these underlying technologies. Primary goals of QoS include dedicated bandwidth, controlled jitter and latency (required by some real-time and interactive traffic), and improved loss characteristics. QoS technologies provide the elemental building blocks that will be used for future business applications in campus, WAN and service provider networks. Almost any network can take advantage of QoS for optimum efficiency, whether it is a small corporate network, an Internet service provider, or an enterprise network. With the development of high-speed networks and multimedia technology, the new applications such as e-learning, video conferencing, telemedicine and so on emerge in an endless stream. The promotion of these new applications depend largely on its ability to meet the practical requirements of relevant service indicators, like that whether the delay increases overlarge, whether the image screen jitter and whether the voice and image is synchronized and so on, these requirements are the so-called QoS(Quality of Service). In order to achieve the QoS requirements, each layer of the network as well as all variety of devices on the network need to work in parallel. As an underlying transmission and switching technology, how does the OBS to provide QoS guarantees like the upper lawyers? It is not only the need to enhance and last the QoS capacity of the upper network, but also the important feature of future optical Internet. Accordingly, the research on OBS networks gets more and more attention, and gradually became the hot research topic in the OBS field.

## 2. AGGREGATION ALGORITHMS COMMONLY USED IN EDGE NODES

### 2.1 Fixed Assembly Period Algorithm (FAP)

Fixed Assembly Period algorithm (FAP)[1] is a very simple and very intuitive aggregation algorithm, whose basic principle is when the maximum delay in queued packet (i.e. the first packet delay) reach assembly threshold, the process of assembly complete, resulting in a BDP (Burst Data Packet) and generating the corresponding BCP (Burst Control Packet). However, the disadvantage of this algorithm is that the BDP is produced almost cyclical, which will cause a very high probability of continuous burst, and not conducive to support the multi-level QoS in OBS networks, as well as not conducive to the realization of priority.

### 2.2 Fixed Assembly Size Algorithm (FAS)

The basic idea of Fixed Assembly Size Algorithm (FAS)[2] is to collect the packets that reaching the same destination node and belong to the same priority to the same queue, when the queue length close to the fixed threshold, then resulting in BDP and generating the corresponding BCP. The disadvantages of FAS algorithm is that when the access load is low, the assembly time will be very long, which is likely to exceed the threshold of real-time services.

### 2.3 Max Burst-size Max Assembly Period Algorithm (MSMAP)

Max Burst-size Max Assembly Period Algorithm (MSMAP) [3] is when one queue length gets assembly threshold, or its corresponding maximum delay achieves assembly time threshold, the corresponding BDP and BCP will be generated. Similar to the FAP algorithm, MSMAP algorithm also exist the problem of continuous burst.

### 2.4 Adaptive Assembly Size Algorithm (AAS)

In terms of related access load, Adaptive Assembly Size Algorithm (AAS)[4] automatically adjust the corresponding threshold, and ensure that assembly time will not exceed a fixed assembly time. In order to avoid fast burst speed along with different access operation in algorithms such as FAP, FAS, MSMAP etc., a window is inducted to the AAS algorithm to reduce burstiness of the operation. Increasing or decreasing the volume of operation to a certain extent, the window corresponding up glides or falls. When a queue length or its corresponding maximum delay achieves assembly threshold, BDP and corresponding BCP will be generated. Whether the window up glide, fall or remain unchanged according to the size of BDP after assembly. Typically, the initial *low* Q is the smallest burst length,$Q_{high} = Q_{low} + a \times \Delta a (a > 0)$ $Q_{high}$ and $Q_{low}$ vary between minimum and maximum BDP length. According to access flow conditions, AAS algorithm can change the size of BDP adaptively; therefore it has a good assembly effect.

**2.5 AAS Algorithm Based on Priority**

Adding Class of Service (CoS) domain to Burst Head Packet (BHP), to make it supporting QoS [5]. Assumes that there are M outlet destination addresses in support of N kinds of operation types, then maintain M ×N queues $Q_{ij}$ in edge routers, each queue maintains a timer $T_{ij}$. So this assembly algorithm can be expressed as the following forms

1. When the class is i, the destination address is j, while packet with the length of L arrives: If $Q_{ij}$ is null, the timer $T_{ij}$ starts counting, $L_{ij} = L$; if $L_{ij} + L > L_{max}[i]$, assembling the packet in queue $Q_{ij}$ into a Burst with the class of $C_i$, meanwhile, $T_{ij} = 0, L_{ij} = L$; else joining the packet into the queue $Q_{ij}$;

2. When timer $T_{ij}$ reaches $T_{max}[i]$, assembling the packet in queue $Q_{ij}$ into a Burst with the class of $C_i$, meanwhile $T_{ij} = 0$, $L_{ij} = 0$. According to different operation types, different $T_{max}$ and offset time can be set to achieve QoS mechanism based on offset time. For example, if voice service set Class1, TCP business Class2, then $T_{max}$ can be smaller, while the $T_{max}$ may be larger, meanwhile allocate a smaller offset time for Class1, thus, the loss rate of Class2 is smaller, but having a large time delay; while the Class1 is opposite.

## 3 THE SUGGESTED MODEL

This session discusses the suggested model for the fiber fault detection and recovery. This model is based on the hybrid burst OCS/OBS and the Genetic algorithm Communication starts when a receiver needs data, provided that the transmitter is ready to deliver it. Consider a network shown in figure1. Assume that N10 (destination) node needs a data from N1(Source) node, there should be a dedicated communication path between N1 and N10.By carefully examining the above network we could find that there are a number of dedicated paths between N1 and N10 like [N1 N2 N3 N4 N5 N10], [N1 N6 N7 N5 N10], [N1 N8 N9 N10], [N1 N6 N10] etc. But the genetic algorithm in the transmitter will find the optimum shortest path between the nodes N1 and N10. Once the shortest path is determined and the data is available then transmitter node is ready to deliver data and communication established between nodes N1 and N10. Now the question is what happens if one of the network edges in the shortest path failures to deliver data to the immediate successor node because of fiber crack or damage. In such a situation the communication between N1 and N10 stops and data is lost. So for preventing this data loss we need an efficient fault detection and recovery system. The current technique uses the protection method which means the optical data is duplicated in to two copies and one copy is explicitly transmitted through the backup light path in parallel with the working light path. This method is like a treatment for disease before the actual situation arises and without any probability for the occurrence of disease in future. Since a copy of data is transmitted through the backup light path, we are actually unnecessarily overloading the network capabilities and wasting the bandwidth and increasing the traffic in such backup lightpath. Now what happens if there occurs a failure in this backup lightpath? It requires another secondary backup. So this will increase the total cost and the complexity in the network. So in this suggested model it is proved that the restoration technique is better than the protection system, provided that we have to simplify the constrains mentioned in the drawbacks of restoration system.

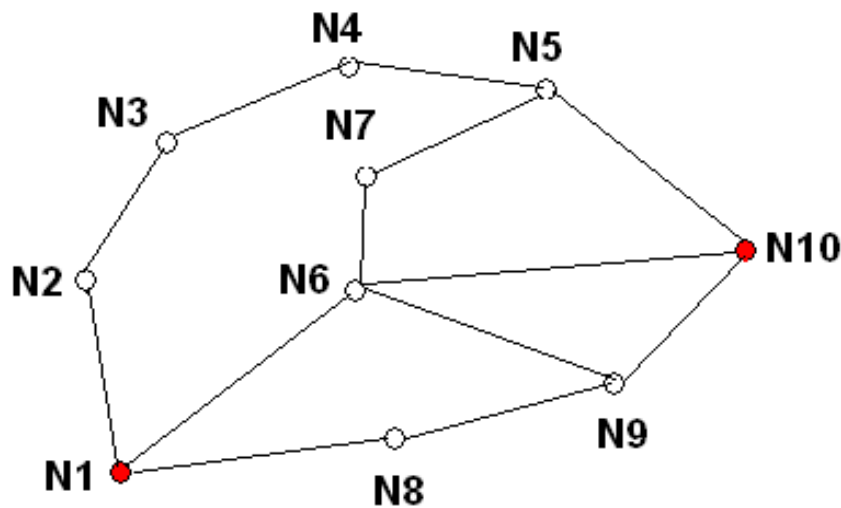

Figure 1. General Network

**3.1 The Fault Detection by Source and Receiver**

The fault detection is vital for both types of the systems. Once the fault is detected, then the source automatically switches the data to the alternate optimum path to the destination. This optimum shortest path is determined by the genetic algorithm.

A. Fault Detection by the Transmitter

At the transmitter there is a small time frame is set for the reception of an acknowledgement signal from the receiver after receiving a burst of data through the reverse channel. Let this time be '$t_s$'. If the transmitter doesn't gets the acknowledgement signal within this time frame then the transmitter realized that there is a fault in the fiber link. Under this situation the transmitter switches the data payload through the alternate optimum path to the immediate successor node. This is achieved by reconfiguring the control packet of each data bursts.

B. Fault Detection by the Receiver

Each receiver node consists of a monitoring unit at the data port and it detects the fiber failure when there is a loss of light. In this model the receiver doesn't want to perform any action and just wait for the data burst to come.

**3.2 Data Switching Technique**

One of the major drawbacks for the recovery system mentioned in the current technology is that both source and destination switches from working lightpath to backup lightpath. But in this suggested recovery system there is no working lightpath and backup lightpath and there is no need for physical switching between these two paths by the source and destination. This is achieved by adopting the Optical Burst Switching (OBS) method. With recent advances in wavelength division multiplexing (WDM) technology, the amount of raw bandwidth available in fiber links has increased by many orders of magnitude.

Meanwhile, the rapid growth of Internet traffic requires high transmission rates beyond a conventional electronic router's capability. Harnessing the huge bandwidth in optical fiber cost-effectively is essential for the development of the next generation optical Internet.

Several approaches have been proposed to take advantage of optical communications and in particular optical switching. One such approach is optical circuit switching based on wavelength routing whereby a lightpath needs to be established using a dedicated wavelength on each link from source to destination. Once the connection is set up, data remains in the optical domain throughout the lightpath. An alternative to optical circuit switching is optical packet switching. In optical packet switching, while the packet header is being processed either all-optically or electronically
after an Optical/Electronic (O/E) conversion at each intermediate node, the data payload must wait in the fiber delay lines and be forwarded later to the next node.

There are two common characteristics among these variants:

- Client data (e.g., IP packets) goes through burst assembly/disassembly (only) at the edge of an OBS network; nevertheless, statistical multiplexing at the burst level can still be achieved in the core of the OBS network.
- Data and control signals are transmitted separately on different channels or wavelengths thus, costly O/E/O conversions are only required on a few control channels instead of a large number of data channels.

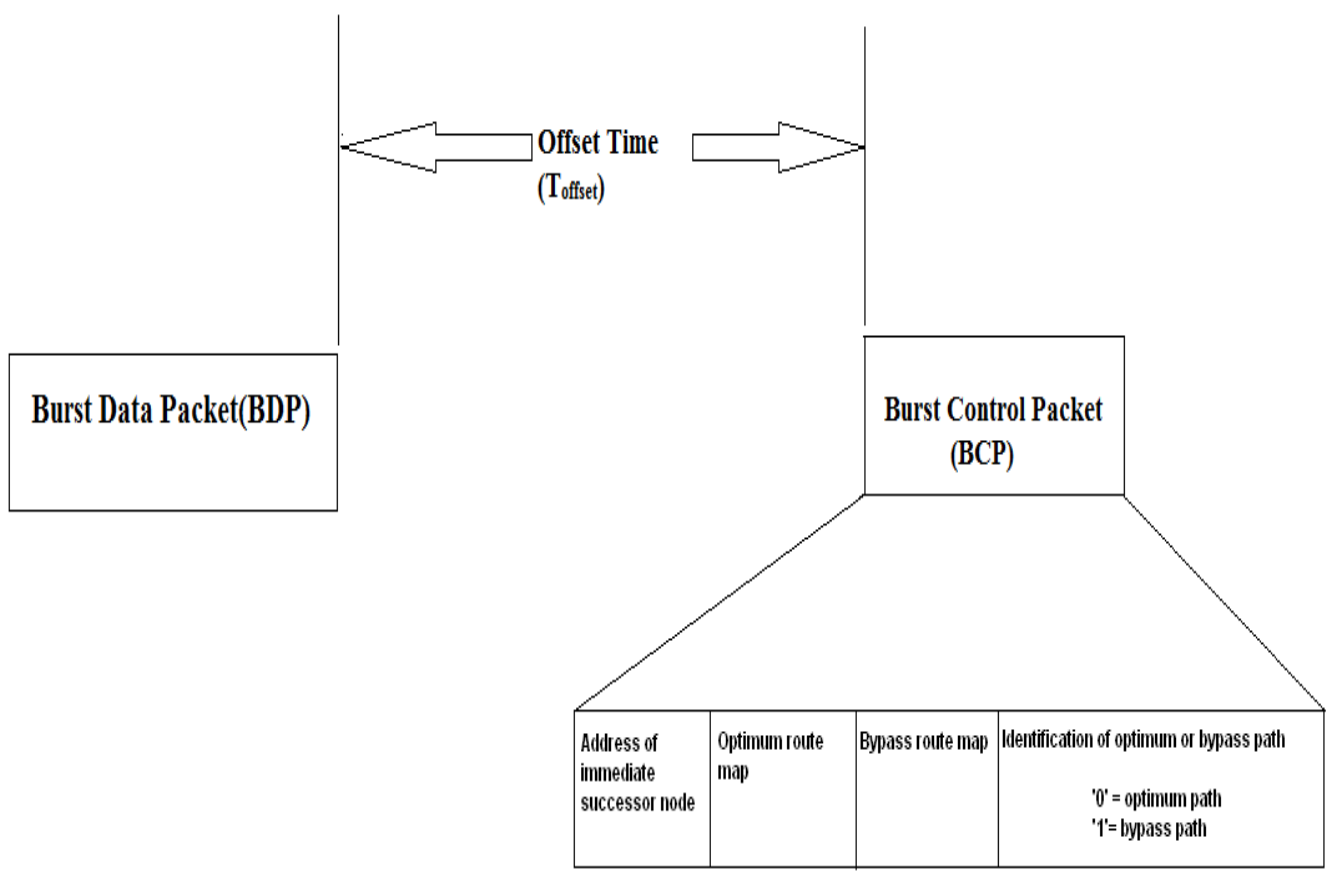

Figure 2. The Burst Structure

The header of the burst consists of four parts, first is the address of the immediate successor, second is optimum route map, this is the actual path determined by the source node towards the remote destination. Third is bypass route map, this is a new route map between the adjacent node whenever there is fault detected in the optimum path. Fourth field is for identifying each node that whether it is in a bypassed path or optimum path. Because the intermediate nodes doesn't know that whether they are in optimum path or bypassed path. The working of each node is mentioned below

Whenever a burst of data is received by the node it will perform the following steps,

1. The header ids processed electronically at each node.

2. It will look in to the last field of the header to identify its membership in the two different paths. If this field is *'1'*, then node understands that it is in the optimum path. If it is *'0'* which means it is in the bypass path. If it reads *'1'*, then the node takes the second field of the header for finding the immediate successor node and this node address is rewrites to the first field. While examining the second field and understands that this is the final destination node then there is no further processing is done on the header and wait for the data payload and terminates

The node performs no change to the last field under two conditions

1. If this node is not the second last node of the bypass path.
2. There is no fault is detected.

If a fault is detected then the node performs three operations,

1. Toggle the last field
2. Determine an optimum bypass path using genetic algorithm and is stored in the third field of the header.
3. And the new successor address is stored to the first field.

If the node reads '0' which means that it is in the bypass node and reads the third field of the header and the first is rewritten by the address of the immediate successor and no change is given to the last field. This is done under the two conditions mentioned above.

For each burst a burst header is transmitted on a common control channel, to be followed by its payload after an offset time on a data channel. The payload passes through the intermediate nodes optically yielding transparency in payload data. The offset time should be taken in to account the header processing time '$t_h$' and the time frame for the detection of acknowledgement from the receiver '$t_c$'

$$T_{offset} = t_h + t_s$$

'$t_h$' consists of the time for checking the fields according to conditions and the time taken for rewriting the header fields

## 4. Conclusion

Optical Burst Switching (OBS) has been developed as an efficient switching technique to exploit the capacity provided by Wavelength Division Multiplexing (WDM) transmission technology for the next generation optical Internet. One critical design issue in OBS is how to provide Quality-of-Service (QoS) in optical networks. So it requires a time effective and less complex methods for maintaining the demanded QoS. This paper demonstrated several latest techniques used for this purpose and finally a new method based on Optical Burst Switching (OBS) and Graph theory is suggested. Since the optical burst switching (OBS) is a promising technique for future optical networks, the suggested model can be a breakthrough in the optical network fault detection and recovery and thus provides an attractive QoS in the optical data transmission and switching.

## Author Biographies

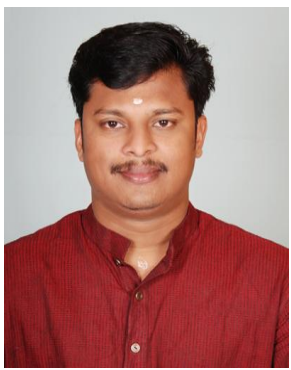

**Sunish Kumar O S**: Mr. Sunish Kumar is currently working as Asst. Professor in an Engineering college at Kerala. He did B Tech in Electronics and Communication Engineering from Cochin University of Science and Technology and MBA in HRM &FINANCE from Kerala University, Trivandrum and now he is pursuing M Tech in Optoelectronics and Communication Systems in CUSAT. Mr. Sunish has authored several papers in National and International Conferences and International Journals. His interested areas are *Communication Systems, Digital Signal Processing, Applications of Fuzzy logic in Management, Optical Burst Switched Networks* etc. Mr. Sunish Kumar is a recipient of Baba Sahib Dr. Ambedkar National Fellowship Award 2011for his contributions to the field of education and cultural activities.